# Graphene Transistor Based on Tunable Dirac-Fermion-Optics

Ke Wang[a,b], Mirza M. Elahi[c], K. M. Masum Habib[c,†], Takashi Taniguchi[d], Kenji Watanabe[d], Avik W. Ghosh[c,e], Gil-Ho Lee[a,f,1], and Philip Kim[a,1]

[a] Department of Physics, Harvard University, Cambridge, MA 02138, US; [b] School of Physics and Astronomy, University of Minnesota, Minneapolis, MN 55455, US; [c] Department of Electrical and Computer Engineering, University of Virginia, Charlottesville, VA 22904, US; [d] National Institute for Materials Science, Namiki 1-1, Tsukuba, Ibaraki 305-0044, Japan; [e] Department of Physics, University of Virginia, Charlottesville, VA 22904, US; [f] Department of Physics, Pohang University of Science and Technology, Pohang, 37673, South Korea

**The linear energy-momentum dispersion, coupled with pseudo-spinors (1), makes graphene an ideal solid-state material platform to realize an electronic device based on Dirac-Fermionic relativistic quantum mechanics. Employing local gate control, several examples of electronic devices based on Dirac fermion dynamics have been demonstrated, including Klein tunneling (2), negative refraction (3-5) and specular Andreev reflection (6, 7). In this work, we present a quantum switch based on analogous Dirac-fermion-optics (DFO), in which the angle dependence of Klein tunneling is explicitly utilized to build tunable collimators and reflectors for the quantum wave function of Dirac fermions. We employ a novel dual-source design with a single flat reflector, which minimizes diffusive edge scattering and suppresses the background incoherent transmission. Our gate-tunable collimator-reflector device design enables measurement of the net DFO contribution in the switching device operation. We measure a full set of transmission coefficients of DFO wavefunction between multiple leads of the device, separating the classical contribution from that of any disorder in the channel. Since the DFO quantum switch demonstrated in this work requires no explicit energy gap, the switching operation is expected to be robust against thermal fluctuations and inhomogeneity length scales comparable to the Fermi wavelength. We find our quantum switch works at an elevated temperature up to 230 K and large bias**

[†] Present address: Intel Corp., Santa Clara, CA 95054 US.
[1] To whom the correspondence should be addressed: pkim@physics.harvard.edu (P.K.) or lghman@postech.ac.kr (G.-H.L.)

**current density up to $10^2$ A/m, over a wide range of carrier densities. The tunable collimator-reflector coupled with the conjugated source electrodes developed in this work provides an additional component to build more efficient DFO electronic devices.**

**Significance Statement**

We build electronic device based on graphene to demonstrate switching device operation utilizing Dirac Fermion Optical analogy for quantum electronic transport. Using collimator and reflector for electron waves, our device shows an order of magnitude improvement compared to previous reported value. We found that the device's electronic characteristics and performance are resilient to environmental change of temperature, bias current, doping and global electrostatic gate offset. The tunable collimator and reflector demonstrated in this work, together with the methods for characterizing their quantum characteristics separately from the trivial contributions, provide the building blocks towards more sophisticated quantum devices. We believe our results will motivate significant interest in applications of relativistic quantum physics and electron-optics for novel device applications.

While the depletion region of conventional semiconducting PN junction blocks the electronic transport across the junction, the gapless band structure of the graphene facilitates electrically adjustable PN junctions and enables novel electronic optics. The transmission probability (*T*) across the PN junction is unity for normal incident electrons due to the pseudo-spin conservation of Dirac fermions (DFs). This startling phenomenon known as Klein tunneling (8, 9) was firstly demonstrated in a graphene PNP junction (2). For the DFs with an oblique incident angle ($\theta$), a PN junction exhibits Snell's law like an electron beam path with a negative refraction medium (3-5) for incoming Dirac electron wavefunctions. However, *T* is exponentially suppressed with $\theta$ as $T \sim \exp[-\pi(k_F(d/2))\sin^2\theta]$ for the symmetric potential of P and N regions, where $k_F$ is Fermi momentum and $d$ is characteristic length scale of potential change across the junction (8, 9). A generalized equation for arbitrary junctions is in Ref. (10). Depending on the value of $k_Fd$, the junction can be transparent or reflective, a result that has been employed for electron waveguiding (11-14), beam splitting (15), Veselago lensing(4), and negative refraction(5) in graphene.

The strong angle dependence of Klein tunneling transmission $T$ has been proposed to realize a new type of switching device based on Dirac fermion optics (DFO) (10, 14, 16-19). Fig. 1a shows a simple device scheme utilizing analogous electron optics. Here, a single layer graphene channel is controlled by several local gates with predetermined shapes, dividing up electron doped (N) and hole doped (P) regions in the channel. The electrons leaving the source electrode pass through the first PN junction orthogonal to the channel direction. This PN junction filters out electrons with an oblique incident angle and collimates electron beams along the channel. The next PN junction, placed at an angle (~45°) blocks the collimated electron beam due to the oblique incidence to the PN junction and reflects it along a path orthogonal to the original. However, in this simplistic device design, the reflected beam hitting the rough physical edge of the device would diffusively scatter (Fig. 1a), leading ultimately to a leakage current into the drain electrode. On top of that, multiple bounces of electrons in between collimator and reflector junctions contribute to the leakage current. To circumvent these diffusive edge scattering and multiple bouncing events, one may design the collimator-reflector to minimize the channel edge scattering. For example, a sawtooth-shaped top gate which can create double reflections sending the incoming DF beam back to source electrode has been theoretically conceived based on DFO (18, 20). However, a recent experimental work employing the saw-tooth local gated DFO device reported a modest on-off switching ratio of only ~ 1.3 (21). This modest device performance was presumably caused by the Klein tunneling at the tips of the saw-tooth gate even at first incidence and the inefficiency of two reflectors in series for subsequent reflections.

In this work, we present a novel dual-source design with a single flat reflector that reduces diffusive scattering at edges and multiple bounces that are otherwise responsible for high off state current leakage.

Fig. 1b shows a schematic diagram of the proposed device and the overall operational procedure. When the central gate region (controlled by gate $V_2$), turns into the opposite carrier polarities of source and drain regions (controlled by gate $V_1$), carriers injected from each source will either reflect back to the same source (oblique incident angle) or travel ballistically to the other source contact (perpendicular incident angle). This collimation-reflection results in suppressed conduction between the source and the drain, and the device is in 'off' state. When $V_1$ and $V_2$ are at the same polarity, the carriers flow ballistically to the drain, and the device is in 'on' state. This device operation scheme has an advantage compared to the aforementioned single source collimator-reflector scheme (Fig. 1a) or a sawtooth-shaped gate structure (18, 20, 21), as there is no significant channel edge contribution and only one reflection can be used for the 'off' operation. Even with a non-ideal reflector, we thus expect considerably enhanced on-off ratio of the switch.

Figure 1b shows electron microscope image of the local gates used for the dual-source device before the integration of graphene channel with two-sources and one-drain electrodes in place. Switching operation of our device can be demonstrated by measuring two terminal resistance $R_T$ between the drain electrode (1) and source electrodes (2&3). A common bias voltage $V_D$ is applied to the source electrodes while the drain electrode is grounded. Two gate regions, collimation gates and the central gate, are controlled by applied gate voltages $V_1$ and $V_2$, respectively. Figure 1c shows the measured $R_T$ as a function of $V_1$ and $V_2$. The resistance map in ($V_1$, $V_2$) plane can be divided into four quadrants separated by the peak region of $R_T$ ~ 8 kΩ, corresponding to the charge neutral Dirac point, $V_1$, $V_2$ ~ 0. These four distinctive quadrants represent the source collimation/central gate/drain collimation regions in the NNN, NPN, PPP and PNP regimes, respectively. We note that the NNN regime has the lowest resistance $R_T$ ~ 500 Ω, while the PPP regime exhibits considerably larger resistance ~ 1.5 kΩ. In an ideal device, we expect a P/N symmetry in the device gate operation due to the particle-hole symmetry in the graphene band structure. However, the graphene channel can exhibit asymmetry in contact resistance due to the metal-induced contact doping (22), which prefers N channel to have lower contact resistance in our devices. The best device performance, therefore, is shown along the PNP-NNN regime, because there arise additional angled PN junctions between contacts and graphene in PNP (off) regime. Figure 1d shows a slice cut of $R_T$ along $V_1$ at a fixed $V_2$ = 5 V, crossing the PNP (off) to NNN (on) regimes. To benchmark our experimental data, we perform semiclassical ray tracing simulation (5) utilizing a billiard model (23-25) coupled with analytical Klein tunnelling equations at junctions (simulation details in Method). For Fig. 1d, channel resistance ($R_{ch}$) is calculated from simulation and $R_C$ (contact resistance) is calculated from Fig. 1c diagonal elements ($V_1=V_2$) (as for every $V_1$ contact resistance is changing). Then total resistance, $R_T=2R_C+R_{ch}$. To fit the off state (P/N/P), we include a random scattering angle around a specular trajectory (following a Gaussian distribution with standard deviation $\sigma_e=15°$) at the edges. Our analysis shows that on-off ratio degrades with increasing $\sigma_e$ as it creates more and more states inside the transport gap. We observe that the off-resistance in the PNP regime is about 6 times larger than on-resistance in NNN regime. The on-off ratio of 6 demonstrated in our device is relatively small to be used as a practical switching device operation at this point. However, it is much larger than previously reported DFO based devices (21) and may in fact suffice for analog applications (19). We also emphasize that the switching operation based on our DFO does not require a 'band-gap' in the channel material, since the device operation relies on Klein tunneling of Dirac fermions which in turn keeps the high mobility of graphene intact in the on state and uses a gate tunable transport gap for off state.

In order to realize complete collimation-filter DFO switch, alignment between the collimated beam and the reflected beam is necessary. Random scatterers in the channel can alter the propagation direction of the beam after collimation, directing beams with wrong incident angles to the reflector. Disorder will thus reduce the filtering efficiency of the collimator-reflector pair. We follow analysis similar to Ref. (26) to probe the disorder induced degradation of collimation-filter DFO switch. We first assign the resistance of single PN junction $R_J$ in the diffusive transport limit, by summing over all incident angles to the junction (8). For completely diffusive transport, we can write the total resistance of the device as a sum of serially connected local resistances, including the contributions from the junction, contact and graphene channels.

In a ballistic graphene channel where the DFO collimation-filter switching is effective, we then expect the measured $R_T$ to be substantially larger than the sum of all the local resistance contributions (26).

We emphasize that the trivial PN junction resistances themselves contribute to the on-off as well, therefore it is important to isolate the DFO contribution from $R_T$. Independent control of carrier density in each gate region in our device design allows us to estimate the resistance contributions from the collimator junctions ($R_{J,1}$) and the reflector junction ($R_{J,2}$) by using different gating schemes (see SI Appendix, Fig. S1). As shown in Fig. 2, the collimation junction governs $R_T$ in the gate configuration A, while the reflector junction governs $R_T$ in the configuration B. Thus, $R_{J,1}$ and $R_{J,2}$ can be probed independently. In configuration A, the beam from the source crosses only the collimator junction before reaching the drain electrode. In this configuration, $R_T$ is expressed as $R_T(V_1,V_2) = R_{C,1}(V_1) + R_{G,1}(V_1) + R_{J,1}(V_1,V_2) + R_{G,2}(V_2) + R_{C,2}(V_2)$, where $R_{C,1}$ and $R_{C,2}$ represent the contact resistance of source and drain electrodes, respectively, and $R_{G,1}$ and $R_{G,2}$ do the graphene channel resistance of blue and green regions, respectively. Here, $R_{J,1}$ is symmetric with exchanging $V_1$ and $V_2$, $R_{J,1}(V_1,V_2) = R_{J,1}(V_2,V_1)$, and vanishes when $V_1 = V_2$, i.e., $R_{J,1}(V_1,V_1) = 0$. As a result, $R_{J,1}$ can be expressed in terms of $R_T$: $R_{J,1}(V_1,V_2) = [R_T(V_1,V_2) + R_T(V_2,V_1) - R_T(V_1,V_1) - R_T(V_2,V_2)]/2$. Note that in this expression, all the terms of $R_C$'s and $R_G$'s are cancelled out and $R_{J,1}$ can be obtained from the measured $R_T$ map. Similarly, $R_{J,2}$ can be extracted from configuration B, where the beam goes through only the reflector junction in front of the drain electrode. Now when the beam from the source goes through both collimator and reflector junctions, we need to introduce an effective resistance $R_{SJ}(V_1,V_2)$ which describes the effect of the collimation-filtering. $R_T$ in this situation (configuration C) can be written as $R_T(V_1,V_2) = R_{C,1}(V_1) + R_{G,1}(V_1) + R_{SJ}(V_1,V_2) + R_{G,2}(V_2) + R_{C,2}(V_1)$. Here, considering the electron-hole symmetry of the graphene channel, $R_2(V_2) \approx R_2(-V_2)$. We also assume a negligible series junction resistance in the unipolar regime, i.e., $R_{SJ}(V_1,V_2) \approx 0$ for $V_1 \cdot V_2 > 0$. Note that $R_{SJ}(V_1,V_2)$ becomes precisely zero for $V_1 = V_2$. Summing up, we can rewrite $R_{SJ}$ in terms of $R_T$: $R_{SJ}(V_1,V_2) = R_T(V_1,V_2) - R_T(V_1,-V_2)$ for $V_1 \cdot V_2 < 0$.

Figure 2 shows $R_{J,1}$, $R_{J,2}$ and $R_{SJ}$ as a function of $V_1$ with a fixed voltage at $V_2 = -5$ V. The finite $R_{J,1}$ and $R_{J,2}$ in the bipolar regime ($V_1 > 0$) is a consequence of the reflected electrons at the PN junctions, whereas small value of $R_J$ for $V_1 < 0$ indicates that the junctions are transparent in the unipolar regime. We also plot $R_{J,1}+R_{J,2}$ to compare with $R_{SJ}$. As we discussed above, if the DFO contribution exists, $R_T$ would be larger than $R_{J,1}+R_{J,2}$. Indeed, as shown in Fig. 2, $R_{SJ}$ is larger than $R_{J,1}+R_{J,2}$ for $V_1 > 0.7$ V, where the two PN junctions are well developed. $R_{SJ}$ is larger than $R_{J,1}+R_{J,2}$ for PNP regime as well (the inset of Fig. 2), directly confirming the DFO switching occurs at both polarities.

We further investigate the DFO switching quantitatively, by analyzing a full set of transmission coefficients $T_{ij}$ between the i-th and j-th terminal in our device as a function of two gate voltages ($V_1,V_2$). Note that the *i* and *j* indices can represent all three electrodes including two source and one drain electrodes. We employ a scattering matrix model in conjunction with the Landauer-Buttiker formalism to compute currents in all possible source-drain and gate configurations to determine $T_{ij}$ (see SI Appendix, Figs. S2 and S3). Figure 3a shows $T_{ij}$ as a function of $V_1$ and $V_2$. The diagonal matrix elements (i=j) represent the fraction of carriers reflected back to the same electrode from which they were injected. In the absence of PN junctions (along with the diagonal line for $V_1=V_2$), the main contribution to the diagonal element $T_{ii}$ represents the probability of carriers being reflected right back at the contact interface in their unsuccessful attempts of getting through. Therefore, 1-$T_{ii}$ is the contact transparency for the $i_{th}$ contact. We find each $T_{ii}$ approaches 0.6 in the NNN regime, consistent with the contact transparency of ~0.4 estimated in the two-terminal resistance (see SI Appendix, Fig. S4).

The off-diagonal matrix elements of $T_{ij}$ contain the quality of DFO switching. In particular, in the presence of PN junctions, we expect the $T_{23}=T_{32}$ (source-to-source reflection) is maximized, and $T_{12}$ and $T_{13}$ (source-to-drain transmission) are minimized. In order to quantify the quality of the DFO switching, we define the relative transmission coefficients, $T_R=2T_{23}/(T_{12}+T_{13})$. Fig. 3b shows $T_R$ as a function of $V_1$ and $V_2$. $T_R$ is expected to be larger in the 'off' regime while it becomes smaller in the 'on' regime. A horizontal line cut of $T_R$ map in ($V_1$, $V_2$) plane at $V_2$ = 5V shows the evolution of $T_R$ from the NPN regime to the NNN regime. The contact transparency along this line is kept high (> 0.4) to minimize its influence on $T_R$. In the absence of PN junctions (NNN regime, $V_1$ > 0), $T_R$ is close to 1, and the injected currents from one of the source contacts are equally split towards the other two electrodes. However, when the PN junctions are established, Klein tunneling across the junction establishes a collimation-reflection effect, increasing $T_R$ above the unity. Figure 3b shows that $T_R$ in the fully developed PNP regime can reach up to 1.4, indicating that DFO switching is effective. Near zero gate voltage (charge neutrality point), carrier motions becomes non-ballistic due to the enhanced effect from disordered electron-hole puddles. In this regime, DFO picture breaks down and our method of extracting $T_R$ becomes inaccurate. This leads to the strongly fluctuating values of $T_R$ near $V_1$ = 0V.

Viewed as a transistor, our DFO switching device exhibits modestly low on-off ratio due to the absence of any energy gaps, and therefore, due to the lack of carrier depletion and device insulation. Instead of band gap, we have introduced a transport gap utilizing angle dependent filtering by the collimator-reflector pair. This transport gap is robust against temperature variations or bias voltages with ideal edges for Klein tunnelling. Even in the presence of diffusive edge scattering it turns into a pseudo gap with a non-zero floor. Thus it provides stability of the device against temperature and bias voltages up to pseudo gap range, which depends on gate voltages (16). The critical device parameters that govern the charge transport characteristics, including contact transparencies, Klein-tunneling probability, carrier densities and quantum conductance for the channels are all insensitive to the temperature and applied bias voltage below critical values, presumably set by inelastic scattering processes. In the graphene channel with hBN, we expect such critical energy scale to be ~ 100 meV, due to optical or substrate induced phonons (27, 28). Figure 4 shows the device characteristic with a wide range of bias currents (up to 150 mA) and temperatures (1.8-230 K). We indeed confirm the stability of the device performance over the entire measured range. The small change in $R_T$ near the peak around $V_2$~-1 V is due to the thermally excited electrons and holes across the Dirac point. However, the device characteristics for high values of |$V_2$| are

not affected by operating temperatures up to 230 K and channel current density up to $10^2$ A/m, demonstrating the robustness of DFO process in our device. We also demonstrate that the on-off device performance can be further improved by engineering the geometric shape of gate electrodes and optimizing DFO (see SI Appendix, Fig. S5).

In conclusion, a quantum switch based on DFO has been investigated, utilizing angle-dependence of Klein tunneling to realize optical analogies of the tunable collimator and reflector. Experimental evidence of DFO characteristics has been demonstrated by isolating the Klein tunneling contribution and extracting a full set of transmission coefficients and low bias resistance characteristics benchmarked against numerical simulations. While the reported on-off ratio is limited by the absence of energy gaps, its robustness against temperature and bias fluctuations can be ideal for many potential applications, such as Dirac-fermion interferometers (29, 30) and analog devices (19).

**Methods**

**Sample fabrication.** The local bottom gates were fabricated by electron beam lithography on $SiO_2$ substrate with palladium-gold metallic alloy. Vacuum annealing of the metallic gates produces a surface roughness of ~0.37 nm which was limited by $SiO_2$ substrate roughness. After fabrication of the local gates, a stack of hBN/graphene/hBN van der Waals heterostructure prepared by dry transfer technique(31) was transferred onto the local gates. The flat surface of the local gate ensures spatially uniform electrostatic gating, hence, well-defined straight PN junctions. High contact transparency of electrodes to the graphene is critical for our experiments as opaque contacts with low transparency would hinder electrons to enter or exit electrodes and lower the visibility of the electronic optical phenomena happening in the graphene channel. Here we adopted in-situ etching technique (4, 32, 33) to achieve highly transparent contacts.

**Simulation method.** Semiclassical ray tracing simulation considers electrons as noninteracting point particles with speed $v_F$ and mass $m=(E_F-qV)/v_F^2$ following classical trajectories (Billiard model) (23-25). Here, $v_F$ is the Fermi velocity, $E_F$ is the Fermi energy, and $q$ is the electrical charge. This has been benchmarked against experiments on graphene PN junctions (5). Electrons are injected from the source at random angles, weighted by a cosine distribution (34). Away from PN junctions the electron trajectories are calculated using classical laws of motion. At the junction, we estimate a fraction $T$ of incident electrons that are transmitted and $1-T$ reflected back. To calculate transmission $T(E)$, we consider a generalized expression considering pseudospin conservation for angle dependent transmission across asymmetric PN junctions (10, 17). In this calculation we use split distance between gates $d$ = 60 nm which is consistent with experimental data (~50-80 nm from SEM and AFM images). The contact $i$ to contact $j$ transmission $T_{ij}(=N_j/N_{Total})$ is obtained by counting the number of electrons ($N_j$) reaching the contact $j$ divided by total number of injected electrons ($N_{Total}$) from contact $i$. Then terminal current $I$ is calculated from Landauer-Buttiker formula by summing up the terminal transmissions.

**Acknowledgments**

The experimental work and theoretical analysis were partly supported by INDEX, a funded center of NRI, a Semiconductor Research Corporation (SRC) program sponsored by NERC and NIST. P.K. acknowledges support from ONR Award No. N00014-16-1-2921 and Lloyd Foundation. K.W. acknowledges a partial support from ONR Award No. N00014-15-1-2761. G.-H.L. acknowledges a partial support from the National Research Foundation of Korea(NRF) Grant funded by the Korean Government(No. 2016R1A5A1008184). K.W. and T.T. acknowledge support from the Elemental Strategy Initiative conducted by the MEXT, Japan and the CREST (JPMJCR15F3), JST.

**Author contributions**

K.W., G.-H.L. and P.K. conceived the idea and designed the project. P.K. supervised the project. K.W. and

Figures and captions

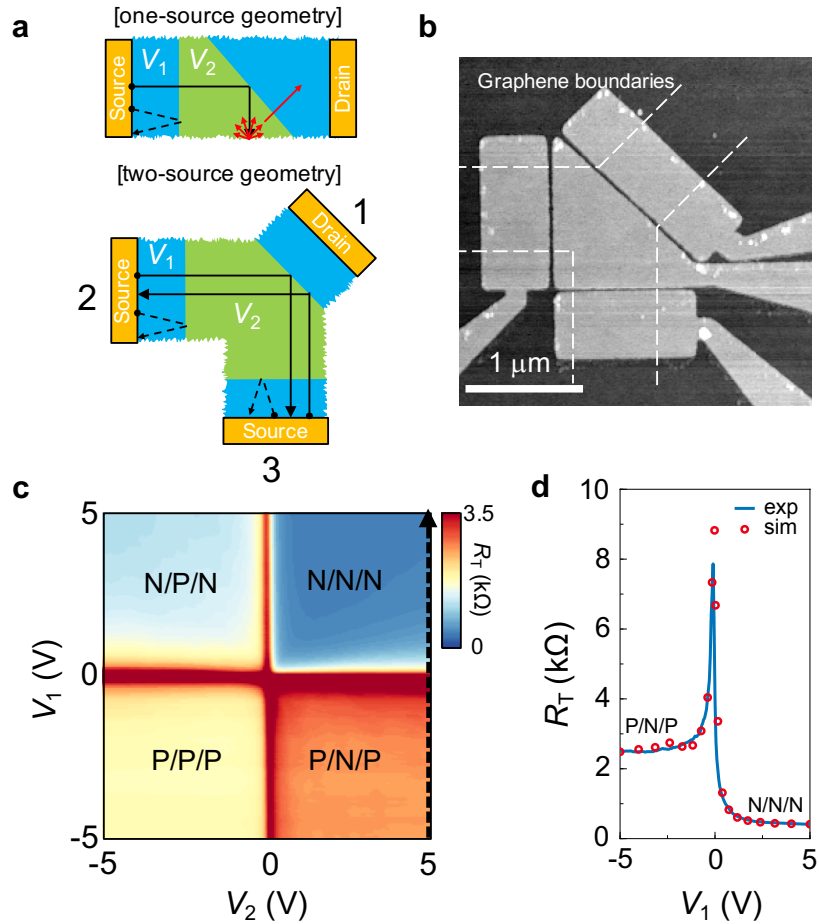

**Figure 1 | Graphene quantum switch. a**, Schematics of the device in the off mode. Central green area (gate voltage, $V_1$) and the blue areas ($V_2$) are doped in different polarity ($V_1 \cdot V_2 < 0$). The collimated electron beams through vertical and horizontal junctions are reflected toward the device edge in one-source geometry or back to the source in two-source geometry. **b**, Atomic force microscope image of bottom gates was taken before transferring a stack of hBN/graphene/hBN. Overlaid broken lines guide the boundaries of graphene. **c**, Color-coded total resistance ($R_T$) as a function of $V_1$ and $V_2$. **d**, Slide cut of the resistance shows the on-off ratio of 6 at fixed $V_2 = 5$ V. Semiclassical ray tracing simulation matches experimental data especially for higher $|V_1|$ (on or off state). To fit the off state (P/N/P), we use edge roughness parameter $\sigma_e = 15°$ (standard deviation of Gaussian distribution of added random angles to specular edge reflections).

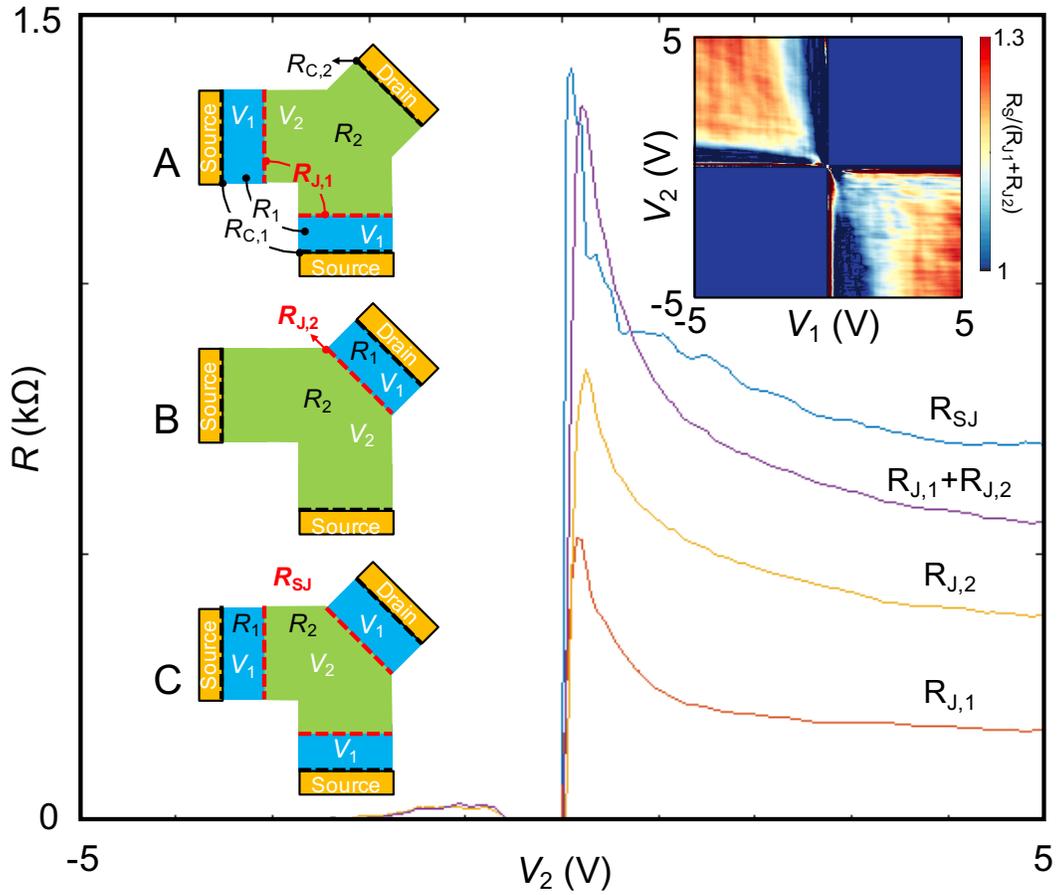

**Figure 2 | Extractions of PN junction resistances.** Resistance contributions from the collimation junctions ($R_{J,1}$), reflection junction ($R_{J,2}$), and series of both junctions ($R_{SJ}$) as a function of $V_2$ at $V_1$=-5 V. The resistance contributions are extracted by symmetrising total resistance to eliminate the contribution from contact and bulk resistances. When PN junction are established ($V_1 \cdot V_2 < 0$), we find that $R_{SJ}$ is always larger than $R_{J,1} + R_{J,2}$, by an amount that corresponds to the contribution from optical behaviour of charge carriers (collimation + reflection). (Inset) $R_{SJ}/(R_{J,1} + R_{J,2})$ plotted as a function of $V_1$ and $V_2$. As PN junction height becomes higher, the contribution from collimation and reflections increases up to ~30%.

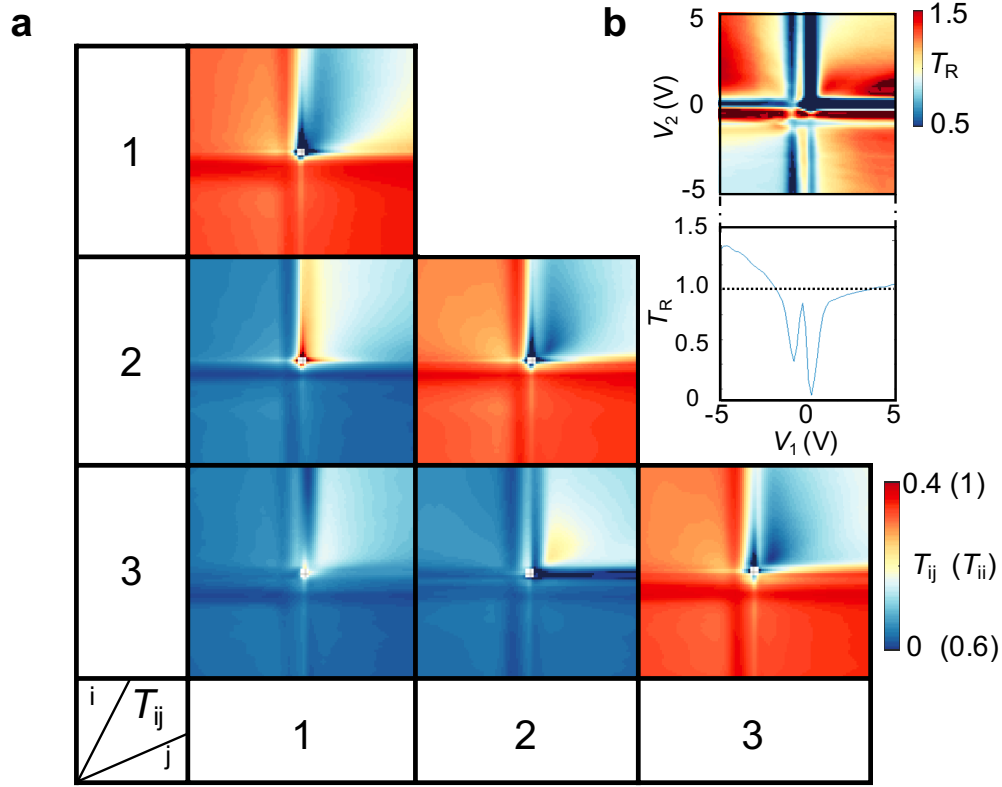

**Figure 3 | Extractions of Transmission Coefficients. a**, Extracted transmission coefficients ($T_{ij}$) as a function of $V_1$ and $V_2$. **b**, Relative transmission coefficients, $T_R = 2T_{23} / (T_{12}+T_{13})$ as a function of $V_1$ and $V_2$, and a 1D cut at $V_2 = 5$ V. In the absence of PN junctions (NNN regime), $T_R$ is very close to 1, and currents injected from any contacts are equally split towards the other two reservoirs. When PN junctions are established, the optical behaviour of Dirac fermions leads to an enhancement of $T_R$ value that is significantly higher than 1 (~1.4). Near zero gate voltage, carrier paths are no longer ballistic due to electron-hole puddles at charge neutrality point. DFO breaks down and renders our method of extracting $T_R$ (SI) inaccurate. This leads to emergence of artifact seen at $V_1 = 0$ V.

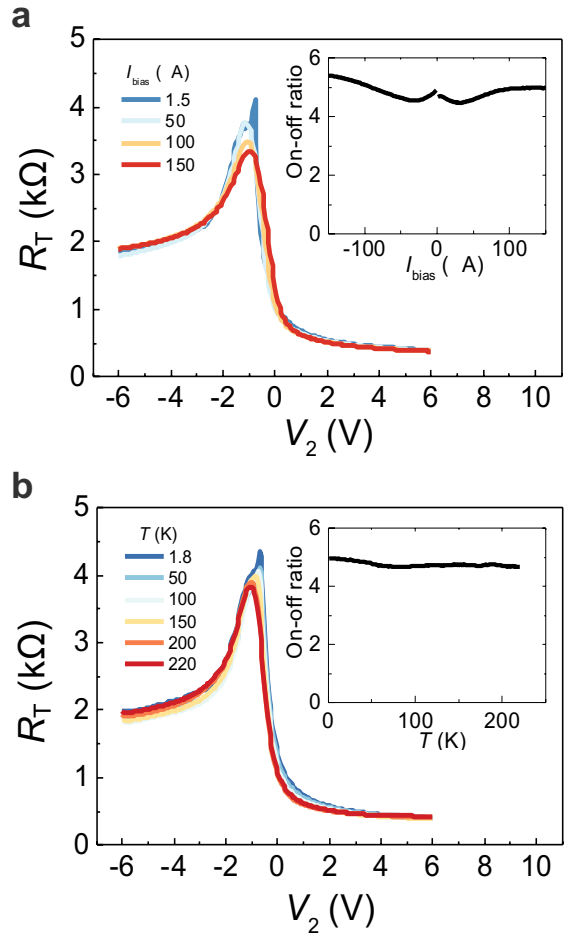

**Figure 4 | Temperature and bias current dependence. a**, On-off behaviour of total resistance ($R_T$) with a central gate ($V_2$) at a fixed collimation gates ($V_1$) at 6 V with various bias current ($I_{bias}$). (Inset) On-off ratio as a function of $I_{bias}$ show no appreciable degradation. **b**, $R_T$ as a function of $V_2$ at $V_1$ = 6 V also shows a robust behaviour against to the temperature ($T$) variation. (Inset) On-off ratio stays the same up to $T$ = 230 K.

Supplementary Information for

# Graphene Transistor Based on Tunable Dirac-Fermion-Optics

Ke Wang[a,b], Mirza M. Elahi[c], K. M. Masum Habib[c,†], Takashi Taniguchi[d], Kenji Watanabe[d],

Avik W. Ghosh[c,e], Gil-Ho Lee[a,f,1], and Philip Kim[a,1]

[a] Department of Physics, Harvard University, Cambridge, MA 02138, US; [b] School of Physics and Astronomy, University of Minnesota, Minneapolis, MN 55455, US; [c] Department of Electrical and Computer Engineering, University of Virginia, Charlottesville, VA 22904, US; [d] National Institute for Materials Science, Namiki 1-1, Tsukuba, Ibaraki 305-0044, Japan; [e] Department of Physics, University of Virginia, Charlottesville, VA 22904, US; [f] Department of Physics, Pohang University of Science and Technology, Pohang, 37673, South Korea

[†] Present address: Intel Corp., Santa Clara, CA 95054 US.

[1] To whom the correspondence should be addressed: pkim@physics.harvard.edu (P. K.) or lghman@postech.ac.kr (G.-H.L.)

**S1. Measuring Individual Junction Resistance.**

In order to isolate the electron-optic process resulted from collimation and reflection at PN junctions, we measure the resistance of PN junctions individually. We first measure $R(V_1, V_2)$, the two probe resistance (using contact 2 and 3 as the source, and contact 1 as the drain) of the device as a function of $V_1$ and $V_2$, in three different gate configurations (Figs. S1a, c and e). We then obtain individual junction resistance (Figs. S1b, d and f) by removing the contribution from contact resistance and bulk graphene resistance via a simple symmetrization process, given by the formula

$$2R_J(V_1, V_2) = R(V_1, V_2) + R(V_2, V_1) - R(V_1, V_1) - R(V_2, V_2),$$

Subtracting individual junction resistance $R_{J1}$ (Fig. S1b, measured with gate configuration Fig. S1a) and $R_{J2}$ (Fig. S1d, measured with gate configuration Fig. S1c) from the combined junction resistance $R_S$ (Fig. S1f, measured with gate configuration Fig. S1e), we obtain the net contribution from the gate-defined electron-optic process, as plotted in Fig. 2 of the main article.

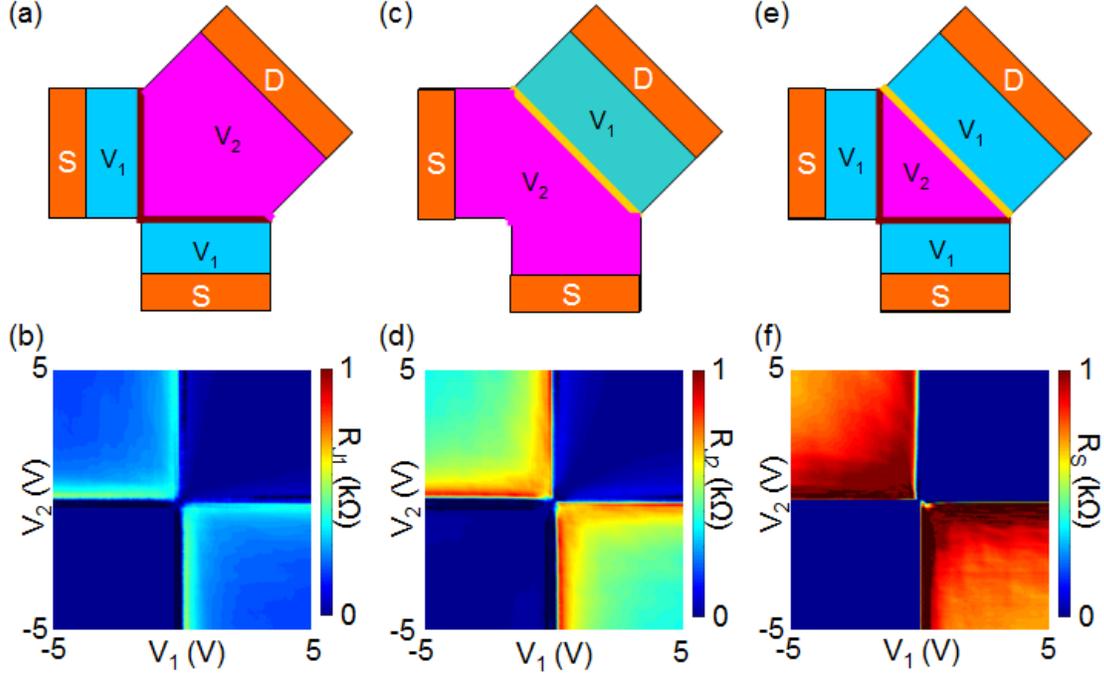

**Figure S1. Measurement of Individual Junction Resistances.** Symmetrized junction resistance, of (a)(b) the collimator PN junctions $R_{J1}$, (c)(d) the reflector PN junction $R_{J2}$ and (e)(f) the combined PN junction $R_S$. $R_S$-$R_{J1}$-$R_{J2}$ yields the non-trivial net contribution from the gate-defined electron-optic process.

**S2. Extraction of Transmission Coefficients.**

The Landauer-Büttiker formalism provides a general description of ballistic transport of multi-terminal structures. In this formalism, all the microscopic scatterings in the conduction channels are summed over and simply represented by a single scattering. At zero temperature, the current $I_i$ in the $i$-th terminal is written by

$$\frac{h}{2e}I_i = \sum_j T_{ij}(\mu_j - \mu_i) = -N_i\mu_i + \sum_j T_{ij}\mu_j, \quad \text{- Eq. 1}$$

where i and j label the terminals and $N_i = \sum_j T_{ij}$ is the total number of modes into terminal i. Basically, this formula describes that the current flowing into terminal i, $I_i$, is the summation of current contributions from other terminals j, $T_{ij}(\mu_j-\mu_i)$, driven by the chemical potential difference, $\mu_j - \mu_i$. Here $T_{ij}$ represents a transmission coefficient from terminal j to i, and the chemical potential is related to the voltage V, $\mu=eV$. Extracting all the components of transmission coefficient $T_{ij}$ provides the full information on the electrical transport of the multi-terminal device. One can consider the simplest way to determine T as shown in Fig. S2. Terminal 1 is biased ($\mu_1 \neq 0$) while other terminals 2,3, and 4 are all grounded to have $\mu_2=\mu_3=\mu_4=0$. Then the current flow into each terminal is measured using Ampere meters. The injected current ($I_1$) from terminal 1 is splitted into other terminals with $I_2$, $I_3$, and $I_4$, that directly determine $T_{21}$, $T_{31}$, and $T_{41}$ as

$$T_{i1} = \frac{h}{2e} \frac{I_i}{\mu_1} \quad (i \neq 1). \quad \text{- Eq. 2}$$

Other transmission coefficients can be measured by changing source terminal from 1 to others and repeating the same procedure. In the experimental situation, however, there are always an additional contribution to the actual chemical potential of the grounded drain electrodes due to the finite resistance from terminals to the actual ground point. The major source of the contributions can be contact resistances, the lead resistance including on-chip connections, electrical lines through cryostats, electrical noise filters, and the offset voltage in the virtually grounded current amplifiers. Since those contributions are path-specific and can vary with external magnetic field or temperature, it is practically hard to keep the chemical potentials of draining terminals to be precisely zero.

This technical challenge can be circumvented by monitoring both of $\mu$ and $I$ for all the terminals. By measuring the actual chemical potentials of each terminal, one can adjust the applied chemical potential accordingly to make the difference between the measured chemical potential difference between the draining terminals and the true ground. However, this method requires an elaborated self-consistent adjustment of multiple variable resistances at every data point. A simpler approach, developed in Ref. (S1), involves a set of measurements in all possible different measurement configurations. This method would generate coupled linear equations where $\mu$ and $I$ are known variables while transmission coefficients are unknown ones. Figure S3 shows an example of all possible measurement configurations when terminal 1 is biased. For each configuration, the current ($I_1$, $I_2$ and $I_3$) and the chemical potentials ($\mu_2$ and $\mu_3$) of draining terminals with respect to the sourcing terminal are recorded. Note that we use multi-terminal device geometry to 'monitor' the chemical potential of the terminal close to the current injection or grounded electrodes. Since there is no current into the voltage probe, the measured chemical potential contains only the negligence contributions from the small finite resistance between the voltage and current probes in the device.

For example, the current through terminal 2 ($I_2$) in the three different configurations in Fig. S3 can be written as

$$I_2^{(\alpha)} = \frac{2e^2}{h}\left(T'_{22}V_2^{(\alpha)} + T'_{23}V_3^{(\alpha)}\right), \quad \text{- Eq. 3}$$

where $T'_{ij} = T_{ij} - N_i\delta_{ij}$ is a redefined transmission coefficient, and each configuration is denoted by the index $\alpha$ =1, 2 and 3, say, for left, middle, and right panels in Fig. S3, respectively. Again, $V_2$ and $V_3$ are the voltages of terminal 2 and 3 with respect to the voltage of sourcing terminal 1, $V_1$, which we set to be zero. These three linear equations can be written in a matrix form as

$$\begin{pmatrix} I_2^{(1)} \\ I_2^{(2)} \\ I_2^{(3)} \end{pmatrix} = \frac{2e^2}{h} \begin{pmatrix} V_2^{(1)} & V_3^{(1)} \\ V_2^{(2)} & V_3^{(2)} \\ V_2^{(3)} & V_3^{(3)} \end{pmatrix} \begin{pmatrix} T'_{22} \\ T'_{23} \end{pmatrix}, \quad \text{- Eq. 4}$$

$$\mathbf{I}_2 = \frac{2e^2}{h}\mathbf{V}\mathbf{T}'_2. \quad \text{- Eq. 5}$$

With linear algebra, the least-square solution for $\mathbf{T}'_2$ can be calculated as

$$\overline{\mathbf{T}'_2} = \left(\mathbf{V}^T\mathbf{V}\right)^{-1}\mathbf{V}^T\mathbf{I}_2. \quad \text{– Eq. 6}$$

After the similar algebraic procedure can be repeated for obtaining other components of $\mathbf{T}'$. $T'_{12}$ and $T'_{13}$ can be obtained by considering $I_1$, while $T'_{32}$ and $T'_{33}$ can be obtained by considering $I_3$. Remaining column of $\mathbf{T}'$ ($T'_{11}$, $T'_{21}$ and $T'_{31}$) can be obtained using the data set taken by sourcing terminals other than terminal 1. Note that this method overdetermines transmission coefficients. In practice, we can use this overdetermined condition to check the reliability of the extracted transmission coefficients.

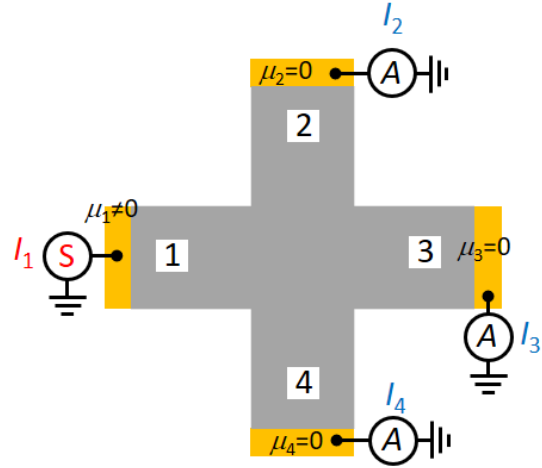

**Figure S2. Extracting transmission coefficients of multi-terminal device.** While terminal 1 is biased (represented by a red letter 'S' in a circle) with chemical potential $\mu_1 \neq 0$ and the rest of terminals are grounded having zero-potential ($\mu_2=\mu_3=\mu_4=0$), draining currents are measured using Ampere meters (represented by a black letter 'A' in a circle).

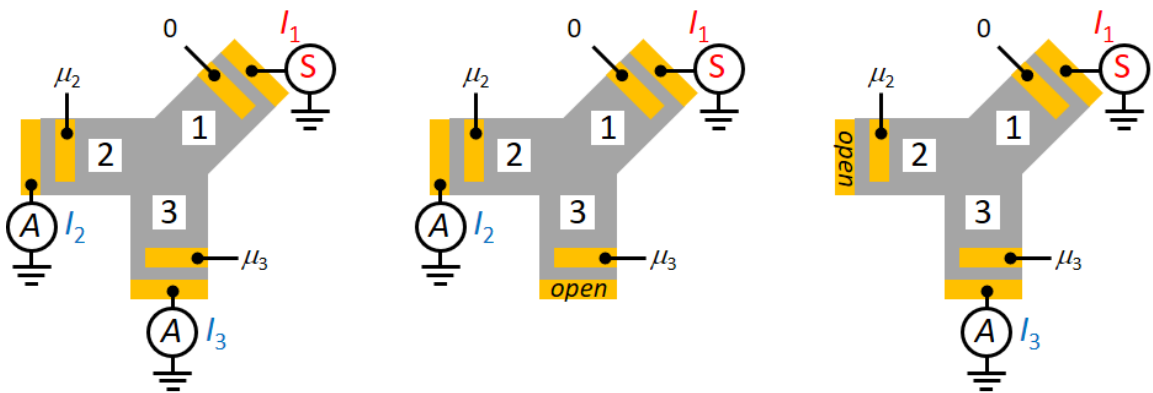

**Figure S3. Three different measurement configurations for three-terminal device with sourcing terminal 1.** Current contacts and voltage contacts are separated for each terminal. While terminal 1 is biased

(represented by a red letter 'S' in a circle), there are three different grounding configurations; grounding both terminals 2 and 3 (left panel), only terminal 2 (middle panel), or only terminal 3 (right panel).

**S3. Achieving Optimum Device Performance.**

While ballistic graphene is not difficult to obtain with proper hBN-encapsulation, in this particular quantum transistor (Fig. S4a), two additional fabrication challenges must be overcome in order to achieve optimum device performance. First, the contact transparency ($T$, defined as the transmission probability of carriers across the contact) needs to be improved in order to maximize the device conductance in ON state, as well as the yield of carrier reception and carrier injection at the source contacts in OFF state. We first lithographically define a mask with standard PMMA electron beam (e-beam) lithography resist and expose a 1D graphene channel for electrical contacts using a reactive ion etching process. We then immediately deposit Cr/Pd/Au to make contact using the same mask. This allows clean graphene-contact interface and minimum gating effect by contact metal. The contact transparency as a function of carrier density (inset) demonstrates high quality n-type contact in our device. Second, it is not a trivial task to integrate local gating capability without compromising the ballistic nature of the graphene. Local gate patterns can introduce strain, remote scatterer and distortion to graphene channel, rendering the device diffusive and inhomogeneous. In order to overcome this challenge, we employ improved lithography by proximity correction, PdAu local gate deposition, and vacuum annealing at 350 ℃. This allows us to fabricate atomically-flat local gates, as shown in the atomic force microscope scan along the purple line in Fig. S4a This device fabrication method enables the integration of local gating capability and ballistic encapsulated graphene in a strain-free and defect-free manner, without compromising the ballistic nature of the device.

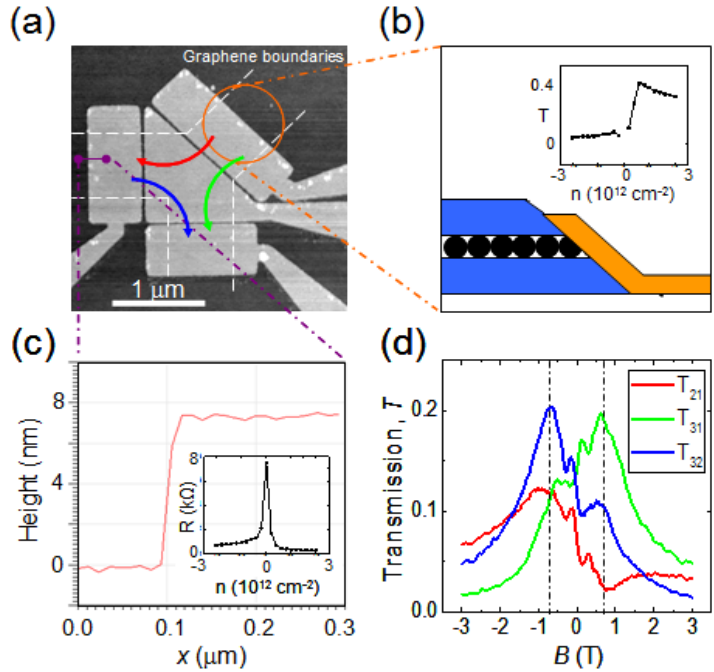

**Figure S4. Achieving Optimum Device Performance**. (a) SEM image of local bottom gates. Two fabrication challenges must be overcome in order to obtain optimum device performance. (b) To achieve transparent ohmic contact (orange cycled region in Fig. S4a), we use the same lithographically-defined mask for etching and metal deposition. This "in-situ" contact method allows clean graphene-contact interface and minimum gating effect of contact metal. The contact transparency ($T$, defined as the transmission probability of carriers across the contact) as a function of carrier density (inset) demonstrate high quality n-type contact. (c) The improved lithography technique, PdAu local gate deposition, and vacuum annealing also allowed us to fabricate atomically-flat local gates, as shown in the atomic force microscope scan (along with the purple line in Fig. S4a). This enables the integration of local gating capability and ballistic encapsulated graphene in a strain-free and defect-free manner, without compromising the ballistic nature of the device. (Inset) The device resistance as a function of carrier density provides a lower-bound estimation of mobility $\mu \sim 200{,}000$ cm$^2$/Vs, and a mean-free path exceeding device geometry (> 2µm). (d) The ballistic nature of the device can also be demonstrated by measuring the magnetic field dependence of transmission coefficient (direction labeled by arrowed-solid-lines in Fig. S4a). The peak positions agree with cyclotron orbit expected in a ballistic device.

The resistance of the resulting device as a function of carrier density (Inset) provides a lower-bound estimation of mobility μ ~ 200,000 cm$^2$/Vs, and a mean-free path exceeding device geometry (> 2μm). The ballistic nature of the device can also be demonstrated by measuring the magnetic field dependence of the transmission coefficient, along with the directions of arrowed-solid-lines in Fig. S4a. The peak positions agree with the cyclotron orbit expected in a ballistic device, thus confirming the ultra-high quality of our local-gated device.

**S4. Offset-Saw-Tooth Device.**

An alternative way of device design to remedy the diffusive scattering in the saw-tooth device is to capture the leakage current with another set of sawtooth gate-pattern, offset laterally to the first set of the sawtooth gate. Figure S5a shows the basic device operation principle. The PN junctions closest to the contacts are perpendicular to the current path, serving as collimator. The reflector is designed as saw-tooth shaped, which prevents carriers from going across the device in OFF state. In addition to the double-reflection process (solid line in Fig. S5a), the double-refraction process (dashed line in Fig. S5a) also contributes to the OFF resistance. Due to our e-beam lithography resolution and also to the finite Fermi wavelength of the charge carriers, both estimated on the order of ~10nm, the tips of these saw-tooth patterns becomes transparent for perpendicularly injected current. Therefore, we added another set of saw-tooth pattern, offset with respect to the original one, to reduce this current leakage in OFF states (dotted line in Fig. S5a). The resulting device yields an improved on/off ratio (Fig. S5d) of ~13. We also conduct a controlled study with devices containing different numbers of saw-tooth gates (Fig. S5b and c), fabricated from the same hBN/Graphene/hBN sandwich. The temperature dependence of the device performance (Fig. S5e) demonstrates no significant difference. This implies that the collimation and reflection happen predominately in the first set of collimators and offset reflectors with high device functionality by design.

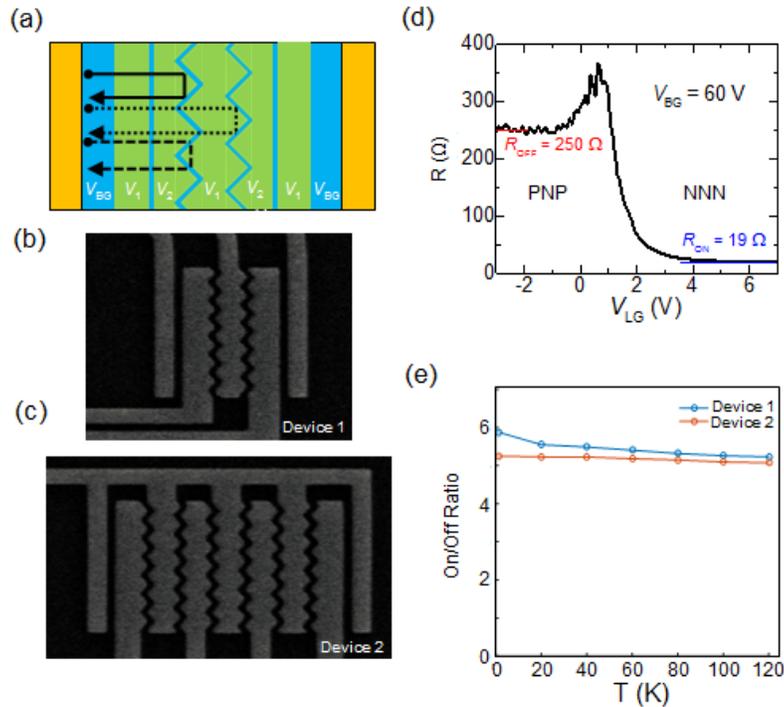

**Figure S5. Improved Device Design with Double Offset Sawtooth Device.** (a) The operation of the device is based on collimator PN junctions perpendicular to the current path located closest to the contacts and saw-tooth shaped reflectors. In addition to the double-reflection process (solid line), the double-refraction process (dashed line) also contributes to the OFF resistance. In addition, the added set of saw-tooth pattern, offset with the original one, reduces the current leakage in OFF states (dotted line in Fig. S5a). (b) This results in an improved on/off ratio off ~ 13. (e) The performance of (b)(c) two control devices containing different numbers of local saw-tooth gates, fabricated from the same piece of graphene. The result demonstrates expected robustness of performance against temperature change, as well as the number of reflectors. This implies efficient collimation and reflection/refraction by design, predominately occurring at the first sets of PN junctions closest to the contacts.

**Supplementary reference**